# Numerical Modeling of Transient Absorption in Hybrid Dual-Plasmonic Au/CuS Nanostructures

Atefeh Habibpourmoghadam*,[1-3], Wenyong Xie[1-3], Patrick Bessel[4,5], André Niebur[1,4], Artsiom Antanovich[6], Dirk Dorfs[1,7,8], Jannika Lauth[1,4,5,6], Antonio Calà Lesina[1-3]

[1] Cluster of Excellence PhoenixD (Photonics, Optics, and Engineering–Innovation Across Disciplines), Hannover, D-30167, Germany
[2] Hannover Centre for Optical Technologies, Leibniz University Hannover, Hannover, D-30167, Germany
[3] Institute for Transport and Automation Technology, Leibniz University Hannover, Garbsen, D-30823, Germany
[4] Institute of Physical Chemistry and Electrochemistry, Leibniz Universität Hannover, D-30167, Hannover, Germany.
[5] Laboratory of Nano and Quantum Engineering, Leibniz Universität Hannover, D-30167, Hannover, Germany
[6] Institute of Physical and Theoretical Chemistry, University of Tübingen Auf der Morgenstelle 18, D-72076, Tübingen, Germany
[7] Institute of Physical Chemistry, University of Hamburg, D-20146, Hamburg, Germany
[8] Cluster of Excellence CUI: Advanced Imaging of Matter, Hamburg, Germany

*E-mail: atefeh.habibpoor@hot.uni-hannover.de

**Abstract**: Transient absorption in plasmonic materials has recently attracted attention of the chemistry and optics communities as a technique to understand dynamic processes and hot carriers generation on ultrafast timescales. In this context, hybrid Au/CuS nanostructures were recently investigated via ultrafast pump-probe transient absorption spectroscopy revealing an exotic dual-plasmonic behavior. Namely, the excitation of a localized surface plasmon resonance (LSPR) in Au (pump at 551 nm) or CuS (pump at 1051 nm), leads to a transient response in the counterpart. This phenomenon was attributed to Landau damping, which stems from hot carrier generation and injection mechanisms at the interface between the two materials. Here, we employ numerical modeling to further clarify the origin of such response in hybrid Au/CuS nanostructures. The geometry of the hybrid nanostructures is first investigated via steady-state simulations (only probe), confirming an UFO-shaped configuration. We provide clarification on the role of the size ratio between Au and CuS. Finally, we present the simulation of transient absorption in the pump-probe regime, which qualitatively replicates our experimental observations, thus identifying the plasmonic response modified via Landau damping as the main governing mechanism. Our numerical approach provides an important tool for the modeling of transient absorption spectroscopy and can support experimental research on dual-plasmonic materials for applications in spectroscopy, photocatalysis, thermoplasmonics, sensing, and energy harvesting.

# 1. Introduction

Hybrid nanostructures (NSs) are formed by two or more nanomaterials which are in touch, in compenetrated, or in close proximity. Usually, NSs made of two plasmonic materials exhibit optical and catalytic properties beyond what is possessed by each single material. Hence, there is great interest in the development of hybrid plasmonic NSs to open up new applications in different fields such as nano-optics, photocatalysis [1],[2], phototherapy [3], surface-enhanced Raman spectroscopy (SERS) [4], and photovoltaics [5]. In colloidal solution, depending on the inter-distance between the nanoparticles (NPs), materials act as isolated objects or show strong resonances due to coupling [6]. Several examples of hybrid NPs have been reported, where the combination of a member of the copper chalcogenide family (e.g., $Cu_{2-x}E$, x > 0, in which E = S, Se, Te) with a plasmonic metal (e.g., Au) have been explored due to their distinctive optical absorption response in the near-infrared (near-IR) and visible spectral regions, respectively [7]. Among all stoichiometric $Cu_{2-x}S$ options, copper sulfide (CuS) offers multiple advantages, such as low toxicity, and high stability against oxidation in air [8]. Bulk CuS is categorized as a heavily p-doped semiconductor, and CuS NSs show plasmonic properties in the near- and mid-IR [13],[14], thus serving as a platform for photoacoustic imaging [9],[10] photothermal therapy (PTT) [11],[12] and tumor treatment [10]. Hybrid NSs formed by combining a member of the copper chalcogenide family with Au are gaining attention for several applications. For example, hybrid Au/CuS NPs can enhance the local accessible electric field distribution in the NPs, which are beneficial for PTT [11],[12]. Moreover, Au/CuS NPs exhibit unique photocatalytic ability due to the electron transfer from the photoexcited CuS to the Au through their common interface [2]. This charge transfer mechanism can further shift the binding energy compared to the single NPs [2] and it can increase the direct ($E_g$=1.7 eV) and indirect band gap energy ($E_g$=1.1 eV) of CuS, when in contact with Au NSs, by a value of ~0.1 eV [2]. In metallic NPs such as Au or heavily doped semiconductors, LSPRs arise from collective oscillation of free carriers at the surface driven by the electromagnetic field of the incident light. For hybrid NSs composed of gold and a copper chalcogenide, in the steady-state regime, the LSPR spectra associated with the two materials are well separated and match the response of the single NPs alone. However, the associated resonances are slightly shifted due to the coupling. In transient absorption spectroscopy (TAS), such hybrid NSs exhibit more exotic phenomena. In Ref. [13], hybrid NSs formed by $Cu_{2-x}Se$ (copper selenide) and Au NPs show a transient response which is regulated by charge transfer mechanisms. This is obtained when the system is pumped with a 100 fs pulse laser at wavelengths of 380 nm and 800 nm, which fall within the interband transitions of Au and $Cu_{2-x}Se$, respectively [13]. In our previous work [15], we excited Au/CuS hybrid NSs at wavelengths matching the LSPR of each constituent, i.e., 551 nm for Au and 1051 nm for CuS. Our characterization showed that a resonant plasmonic excitation of only one material (e.g. Au or CuS) leads to a transient response associated with the other material (either CuS or Au). Resonantly exciting Au/CuS NSs with at 551 nm or 1051 nm gives rise to a plasmonic resonance either in Au or CuS, with following generation of hot carriers and accumulation at the interface between the two materials. This is due to Landau damping (LD), which in turn modifies the optical response of each constitutive material over a thin layer [16]. The process takes place over a time interval shorter than 100 fs. Our previous work showed that 400 fs (for shorter intervals the instrument response function was too long) after initial excitation, the hybrid Au/CuS system exhibited the same transient response, no matter whether the excitation pulse was at 551 nm or 1051 nm. As an important comparison we showed that a simple mixture of CuS and Au particles do not reproduce the behavior and that the pump power dependence was linear, excluding simple multi-photon absorption processes as explanation [15]. Here we focus on better understanding the key mechanism in the ultrafast interaction between the two plasmonic domains in hybrid Au/CuS NPs.

In this paper, we attempt to clarify the physical mechanism regulating the response of hybrid Au/CuS NPs both in the steady-state and transient regime via numerical simulations. Investigations in the steady-state regime were reported in Ref. [7], but a discrepancy between simulation and experimental results still exists. We start by clarifying the geometry of Au/CuS hybrid NSs and take into account the dielectric anisotropy properties of CuS, which leads to an orientational-dependent plasmonic coupling. The identified geometry is ultimately confirmed via transmission electron microscopy (TEM) experiments performed under tilted illumination. In Section 2, we perform numerical simulations to understand the effect of various geometric parameters, such as shapes, interdistance of the center of particles, and spatial orientation of the hybrid NS with respect to the propagation direction of the incident light beam. In Section 3, we focus on steady-state experiments and simulations to further clarify the interplay between Au and CuS NPs. In Section 4, we investigate the optical response of hybrid Au/CuS NSs in the transient regime by performing pump-probe simulations as a function of the delay between pump and probe signals. Our study captures the decay of the plasmonic response and the impact of the hot carriers' mechanism at the early stage of the time-evolution of the system (< 100 fs), before the thermalization process becomes dominant. This is obtained by modeling the Landau damping, which ultimately allows us to qualitatively reproduce the results of the transient absorption (TA) measurements, and gain insight into the role of this physical mechanism.

## 2. UFO-shaped hybrid NSs

Based on the observation of 2D images in scanning electron microscope (SEM) or in TEM, hybrid NSs are often interpreted as core-shell structures [7], which might not be the real case everywhere. Core-shell geometries are extensively explored in literature due to their ability to tailor the optical response based on the hybridization of plasmonic modes [17], as well as for the availability of an analytical solution (Mie theory) [6],[16]. To strengthen the understanding of the Au/CuS NP geometry discussed here, TEM was performed under different tilting angles of the sample. Our investigations confirm that Au/CuS NSs are not in core-shell geometry since the shell does not surround the core completely. A TEM image (not tilted) of drop-casted Au/CuS NPs is shown in Fig. 1a. For most of the observed Au/CuS NSs, CuS NPs have a disk-like shape, and Au NPs are attached to the large side of the disk lacking preferential position. Systems formed by two touching NSs are typically called Janus NPs; we will refer to our Au/CuS NPs as UFO-shaped. We note that the disk-like shape of CuS is also observed in pure CuS NPs [18]. While the optical response of Au/CuS crystalline spherical structures is majorly located in the visible and near-IR, enabling SERS [19], the absorption of UFO-shaped NSs covers a larger wavelength range in the near-IR, which is better suited for biological applications [10].

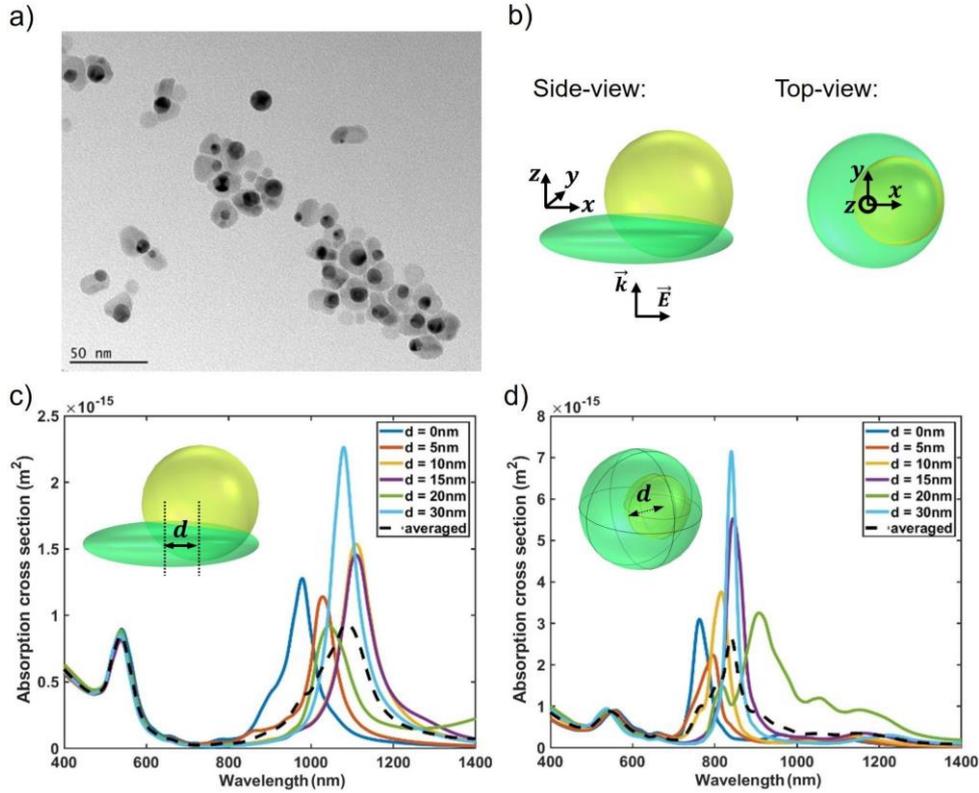

**Fig 1**. a) TEM image of hybrid Au/CuS NPs. b) Schematic representation of the UFO structure. Absorption cross section spectra by progressively increasing the distance *d* between Au and CuS NPs until there is no overlap between the two particles (mixture phase) in the c) UFO-shaped and d) core-shell configuration. The insets show the schematic geometries of the UFO-shaped and core-shell NSs, respectively. The optical response in c) is consistent with experimental results in [15].

To clarify the impact of the geometry on the optical response of the Au/CuS hybrid NS, we present a steady-state study comparing UFO-shaped (as introduced in Fig. 1b) and core-shell [21] NSs. Although all hybrid Au/CuS NPs are surrounded by a ligand shell of oleylamine molecules (with a thickness of 1 - 2 nm) to make colloidal particles stable in toluene, the permittivity of such thin shell is similar to that of toluene. Therefore, toluene is always assumed as the surrounding material. The plasmonic response of Au and CuS can be theoretically and numerically studied by means of classical electromagnetism. For small particles with size well below the incident light wavelength, the application of the quasi-static approximation is rather common [20],[21]. However, the complexity of the Au/CuS NP geometry and material properties require full-wave simulations, which we performed in Comsol Multiphysics (wave-optics module, frequency domain solver). Unless otherwise specified, throughout this paper we model the UFO-shaped structure by considering a Au NP (radius of 10 nm) slightly embedded into a CuS disk particle (radius and thickness are 15 nm and 4 nm, respectively) with an initial interdistance of *d* = 0 nm between the axes of the two particles (see Fig. 1c). The NS is excited by a plane wave propagating in the z-direction and polarized along the x-axis with an amplitude of 1 V/m. The steady-state response is obtained via simulations in the frequency domain over the spectral range of 400-1400 nm. We model Au with the permittivity data in Ref. [22]. The anisotropy of CuS is modeled with permittivity data from Ref. [8]. In the visible spectral range, the real part of both dielectric constants of CuS, namely $\varepsilon_x$ and $\varepsilon_z$, are positive, whereas only $\varepsilon_z$ of CuS keeps the same sign from the visible up to a wavelength near 1051 nm [8], supporting the formation of surface plasmon polaritons (SPPs). This is where the real part of the dielectric constant of Au has always a negative value. Hence, observing a dual-plasmonic response in Au/CuS hybrid NSs is expected, as already reported experimentally [15]. Due to the variation of the interdistance between the two particles in the experimental sample, for both UFO-shaped and core-shell, simulations were done by displacing the center of Au relative to the center of CuS along the x-axis until the two particles are completely separated, hence, leaving no mutual interaction (Fig. 1c,d). In hybrid Au/CuS NSs, the optical response is strongly affected by the anisotropy of CuS and the orientation of its permittivity tensor with respect to the Au NP. Therefore, in UFO-shaped NSs, we model the anisotropy of CuS by considering the c-axis along the z-axis. In fact, for the size range of investigation, the CuS disk NPs preferentially grow laterally in a direction perpendicular to their c-axis [8]. In Fig. 1c, we present the absorption cross section for an UFO-shaped NS as a function of distance $d$, where we observe that the LSPRs associated with Au and CuS are well separated and with similar amplitude, with a spectral shift of the CuS LSPR due to mutual coupling, as reported in experiments [15]. For comparison, in Fig. 1d we consider a core-shell NS formed by Au and CuS nanospheres of radius 10 nm and 15 nm, respectively, where $d$ is the distance between their centers. As $d$ progressively increases, the system evolves from a hybrid NS to a mixture of two distinct NPs ($d$ = 30 nm). In this case, the response of the CuS nanosphere is dominant due to its larger size, and the separated LSPRs are less obvious. In fact, comparing Fig. 1c and Fig. 1d reveals that dual-plasmonic response is less balanced in magnitude in the core-shell configuration than in the UFO-shaped geometry, which we attribute to the enhanced photoexcitation probability of the CuS NS in the core-shell geometry. Moreover, the spectral position of the resonance associated with CuS is blue-shifted compared to the case of a CuS NP alone or the UFO-shaped configuration. Such observations validate the UFO-shaped geometry of our experimental samples.

Since hybrid Au/CuS NPs can have any arbitrary orientation with respect to the incident light when suspended in colloidal solution, we study the absorption cross section for several possible alignments of the UFO-shaped NS. We consider a fixed interdistance of $d$ = 5 nm between the axes of the two particles. The orientation of the UFO-shaped structure can be described based on a unitary vector $n$ along the CuS c-axis. As we rotate the hybrid NS, we also need to rotate the permittivity tensor in order to keep the c-axis properly aligned with respect to the gold NP. In Cartesian coordinates, $n = (n_x, n_y, n_z) = (sin\theta cos\varphi, sin\theta sin\varphi, cos\theta)$, where $\theta$ and $\varphi$ are the tilt angle and azimuth angle, respectively. The azimuth angle $\varphi$ is described in the xy-plane with respect to the x-axis. The tilt angle $\theta$ is described with respect to the z-axis. In general, for $\theta \neq 0$ and $\varphi \neq 0$, the permittivity tensor can be described as $\varepsilon_{ij} = \varepsilon_\perp \delta_{ij} + \varepsilon_a n_i n_j$, where $n_{i,j}$ are the components of the normal vector $n$ as introduced earlier, $\varepsilon_a$ is the dielectric anisotropy defined as $\varepsilon_a = \varepsilon_{||} - \varepsilon_\perp$ and, $\varepsilon_\perp$ and $\varepsilon_{||}$ are the dielectric permittivities perpendicular and parallel to the CuS c-axis, respectively. If $\theta = 0$ and $\varphi = 0$, the complex permittivity tensor can be simplified as $\varepsilon = [\varepsilon_x, 0, 0; 0, \varepsilon_y, 0; 0, 0, \varepsilon_z]$, where $\varepsilon_\perp = \varepsilon_x = \varepsilon_y$ and $\varepsilon_{||} = \varepsilon_z$. The absorption cross section spectra are obtained by rotating the UFO-shaped NS by an angle $\varphi$ in the xy-plane (Fig. 2a), and by tilting the NS by an angle $\theta$ relative to the z-axis (Fig. 2b). In Fig. 2a, by rotating the structure by, e.g., $\varphi = 90°$, the LSPR in the near-IR blue-shifts. The average value of absorption is shown in green color which is centered at λ = 1020 nm. As seen from Fig. 2b, changing the tilt angle $\theta$ can effectively tune the LSPR response in CuS due to the anisotropy while it has negligible impact on the LSPR of the Au particle. It is interesting to observe that the resonance associated with CuS reaches a maximum at around $\theta = 20°$. In general, the orientation of the hybrid NS does not have a remarkable impact on the spectral position of the resonances with a maximum shift of 28 nm in the NIR as a function of $\varphi$. Meanwhile, the random orientation of such NSs in an experimental setup leads to a broadening of the overall optical response, as also evident in Fig. S3.

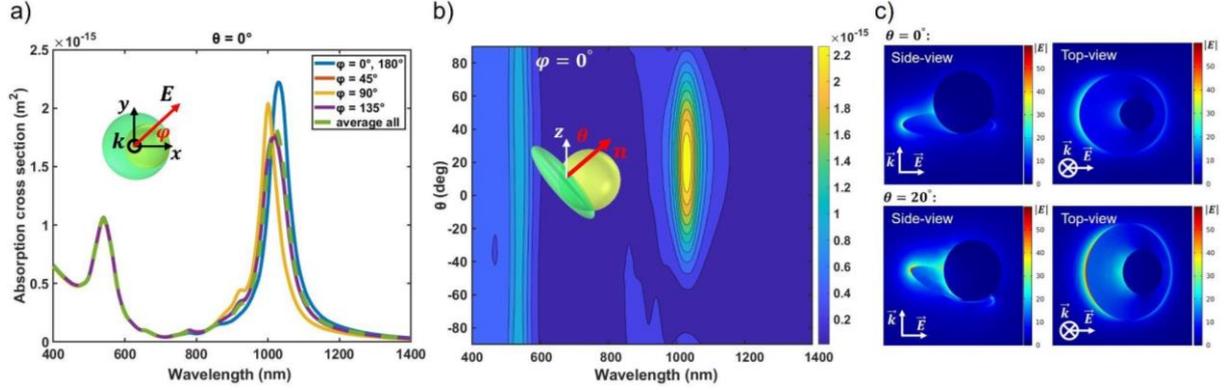

**Fig. 2.** a) Absorption cross section spectra as a function of the UFO-shaped NS orientation with respect to the light polarization (rotation by $\varphi$). b) Absorption cross section of the UFO-shaped NS orientation with respect to the light propagation direction (rotation by $\theta$). For the UFO-shaped geometry, the interdistance between the axes of the two particles is fixed to d = 5 nm. c) Electric field distributions (in V/m) in the UFO-shaped NS in the stationary state for $\theta = 0°$ (top) and $\theta = 20°$ (bottom), respectively. The cut-planes are taken through the middle of the sphere for the side-views and through the middle of the CuS disk for the top-views.

The electric field magnitude distribution (in V/m) for the Au/CuS hybrid NS is shown in Fig. 2c, at the corresponding excitation wavelengths of CuS. At $\theta = 0°$, the plasmonic excitation wavelength of CuS is found to be equal 1051 nm, while at the tilt angle of $\theta = 20°$, this takes place at the wavelength of 1020 nm with some blue-shift. It shows from Fig. 2c, tilting the structure with respect to the incident propagation direction, can excite a resonant response in the CuS disk particle along the z-axis which can further amplify the plasmonic response of the CuS particle in the IR spectral range.

### 3. Size dependence of the dual-plasmonic response in the steady-state regime

Here we focus on steady-state measurements and numerical modeling in order to understand the interplay between the relative size of Au and CuS in hybrid NSs. The dual-plasmonic response of hybrid Au/CuS NSs was studied in the visible [7] and near-IR [13] previously when the size of only one constituent is changed. It was found that the ratio between the magnitude of the two plasmonic resonances can be tuned by controlling the size of Au while keeping the size of CuS constant [7]. In more detail, by reducing the size of the Au NP, the relative plasmonic response of CuS is enhanced in the near-IR region and approaches the response of only CuS. On the other hand, in Au/$Cu_{2-x}$Se samples it was shown that by keeping the size distribution of Au constant, increasing the size of the $Cu_{2-x}Se$ domain results in a significant LSPR enhancement in the near-IR region with minor impact on the response associated with Au in the visible [13].

Here, we aim to clarify the impact of a CuS-domain size variation on the Au response. We perform the synthesis of hybrid Au/CuS NPs with different sizes of the CuS-domain via a seed mediated growth method (more information is provided in the Methods section) and show the TEM images in Fig. 3a-d as well as the corresponding steady-state absorbance in Fig. 3e. Further statistical data is provided in Fig. S1 and Table S1. In the prepared samples, the majority of the obtained NPs exhibit in the hybrid Au/CuS form. A few pure CuS NPs are also obtained. Pure Au NPs are only rarely observed. The mean diameter of the Au/CuS NPs used as seeds in the seed mediated growth reactions is 15.6 nm ± 2.3 nm. With this growth reaction the diameter could be increased to 21.1 nm ± 2.6 nm after one, 26.5 nm ± 3.7 nm after three and 31.1 nm ± 4.9 nm after five growth steps.

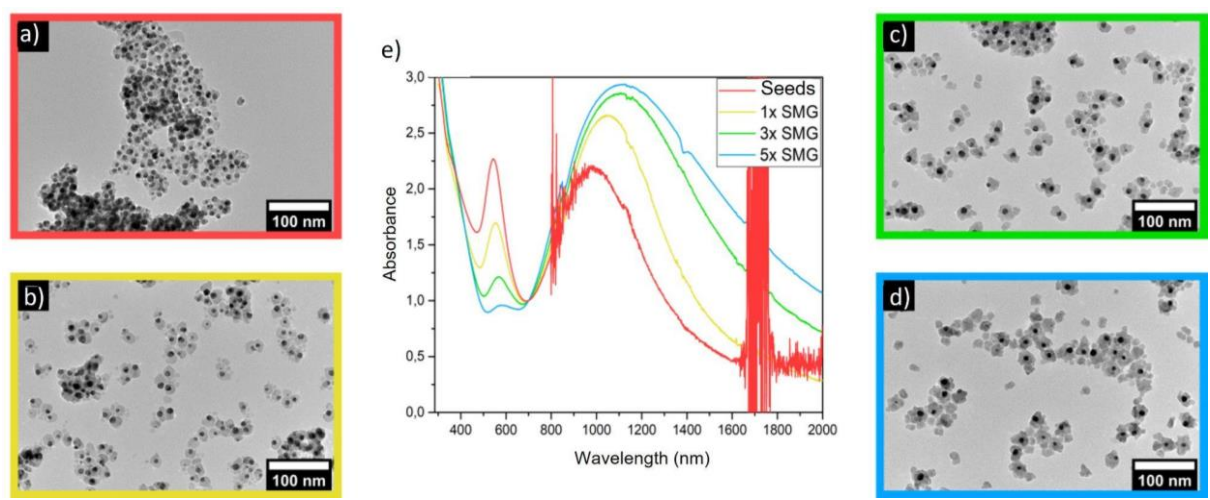

**Fig. 3.** (a-d) TEM images of the hybrid Au/CuS NPs. (a) Hybrid Au/CuS NPs used as seeds in the seed mediated growth (SMG) synthesis. (b-d) Hybrid Au/CuS NPs after one, three and five seed mediated growth steps respectively. (e) UV/Vis/near-IR-spectra of the hybrid Au/CuS NPs shown in the TEM images, normalized to the absorbance at 700 nm.

The steady-state absorbance spectra of the prepared hybrid NPs (Fig. 3e) exhibit two bands in the visible and near-IR regions attributable respectively to Au and CuS LSPR. With the increase of the CuS domain size, the LSPR band of the Au domain shifts towards longer wavelengths and decreases in magnitude relatively, in line with previous observations [15]. This behavior can be hypothesized based on the increase of the CuS-domain and associated increasing of the permittivity of the surrounding medium from $\varepsilon_m(toluene) = 2.22$ [23] to $\varepsilon_{m, 550\,nm}(CuS) \approx 6.4$ [24] (calculated from the refractive index n via $\varepsilon \approx n^2$). An increase of the permittivity of the surrounding of a plasmonic NP is characterized by a red-shift in the LSPR band [25]. The LSPR of the CuS-domain also shows an increasing red-shift with an increasing CuS-domain size. As Au has a lower permittivity than toluene ($\varepsilon_{m, 1050\,nm}(Au) \approx 0.07$), the Au-domain should lead to a blue-shift of the CuS-LSPR. With increasing CuS-domain size however, the influence of the Au-domain is decreased and the LSPR red-shifts to values as they are observed in pure CuS NPs [18].

In order to support the experimental findings we perform simulations of hybrid Au/CuS UFO-shaped NSs with various sizes of the two main constituents. In Fig. 4a, we show a comparison between experimental absorbance (i.e., absorption plus scattering cross section) and simulated absorption cross-section. From simulations we find that the scattering cross section is negligible (a 100-fold smaller) in comparison to the absorption cross section. We consider the same UFO-shaped model already investigated in Section 2. Since in a real setting Au/CuS NPs have random alignment, we make an average over all possible spatial orientations for $-\pi/2 \leq \theta \leq \pi/2$ and $0 \leq \varphi \leq 2\pi$. The experimental spectrum for Au/CuS (see red curve in Fig. 3e corresponding to the seeds case) and the simulated one are in good agreement in the spectral position of their peaks. For comparison, we also show in Fig. 4a the numerical absorption cross section of individual Au and CuS NPs. The Au/CuS spectrum shows two distinct resonances associated with the Au and CuS counterparts. The LSPR associated with the Au-domain is slightly red shifted compared to the resonance of the single Au NP. Meanwhile, the LSPR associated with the CuS-domain shows a blue shift with respect to the resonance of the isolated CuS NP and confirms our previous findings for steady-state ensemble measurements of colloidal Au/CuS solutions [15]. Concerning the relatively broadband absorption of the CuS domain in colloidal solution in comparison to the relatively narrow bands calculated for this absorption band, it should be noted that in addition to geometrical factors of the CuS domain (e.g. the aspect ratio of the disks) and geometrical aspects of the hybrid NS (mainly the position of the Au relatively to the CuS, the parameter "d" in Fig. 1), also the exact chemical composition of the CuS domain can play a crucial role and might also vary within the ensemble of colloidal NPs, therefore causing a further broadening of the near-IR absorption band. Indeed for all investigated copper chalcogenide compounds (e.g. $Cu_{2-x}Se$, $Cu_{2-x}S$ but also $Cu_{1.1}Se$) a drastic dependence of the LSPR frequency on the copper-to-chalcogenide ratio is reported, where tiny variations caused a significant shift of the resonance frequency. In all these cases, a change in the copper-to-chalcogenide stoichiometry directly changes the free charge carrier density in the material and therefore directly changes the plasma frequency of the material [24],[26],[27]. In Fig. S4, we also show the absorption cross section of a pure CuS disk as a function of disk radius and thickness, highlighting the impact of size distributions on the broadening of the overall absorption response in the near-IR spectral range.

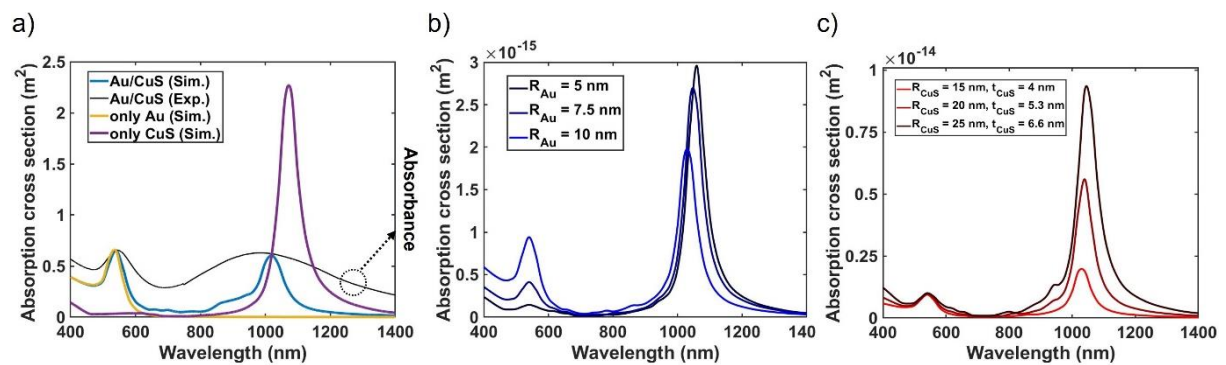

**Fig. 4**. a) Simulated absorption cross section of the Au/CuS UFO-shaped NS compared to the experimental absorbance spectra (corresponding to the seeds sample in Fig. 3e). Absorption cross sections in only Au and only CuS NPs are shown for comparisons. Impact of size variations of b) Au, and c) CuS particles on the dual-plasmonic response of the UFO-shaped NS.

Simulations in Fig. 4b show that by decreasing the Au size (i.e., radius of 10, 7.5 and 5 nm) while keeping the radius of CuS fixed to 15 nm, the plasmonic response of CuS increases and gets close to the single particle response. This can be understood as a result of diminishing the mutual coupling in the system by decreasing the size of Au and thus the common interface of the two NPs. A less pronounced mutual interaction is seen by changing the size of CuS (i.e., radius of 15, 20, 25 nm) while keeping the Au radius fixed to 10 nm. In this case, we observe that increasing the size of CuS, mainly the response in the IR regime is intensified, while the plasmonic response of Au only shows slight variations (Fig. 4c). Here, the thickness of CuS is also increased with the same spatial ratio as its radius. In general, our simulation results confirm the findings previously reported by us assuming the size of one NP fixed and changing the size of the other. In our current experimental setting, the size of the Au NPs is not fixed, but has a statistical distribution with a constant mean average size and a quite large standard deviation (see Figs. 3 and S1). This explains the tunability in the Au-LSPR as a result of increasing the CuS-domain size. We highlight that simulating one single hybrid NS may not be enough to capture all the complexities of the real system, such as the additional coupling between NSs and the cluster topologies that may arise during the growth process. Therefore, considering the statistical size distribution of Au NPs and the impact of seed mediated growth on the geometrical configurations might expand the numerical investigations.

### 4. Transient absorption modeling

In optical pump–probe spectroscopy, hot electrons can be generated in hybrid NSs by an optical pump pulse train. A second broadband (white light continuum) pulse train, the so-called probe, with a typically low energy density is employed to provide information on the photoexcitation states, non-radiative recombinations and decay process of free charge carriers at different time delays with respect to the pump [28]. Femtosecond TAS is employed to study the time-evolution of Au/CuS hybrid NSs after photoexcitation with variable wavelengths, and monitored via a white light continuum from visible to near-IR acting as the probe. Previous TAS analysis for hybrid Au/CuS NSs revealed an exotic interplay between the two plasmonic materials, namely, the LSPR excitation of one material producing a response at the LSPR wavelength of the other material. Excitation wavelengths are equal to 551 nm (2.25 eV) for Au and 1051 nm (1.18 eV) for CuS. In this section, we focus on the numerical modeling of TA. The time varying response of UFO-shaped dual-plasmonic NSs is simulated in Comsol Multiphysics (wave-optics module, transient interface) by numerically solving the instantaneous vector wave equation. To simulate the TA process, we use a 100 fs pump pulse centered at the LSPR wavelength of interest, and a 20 fs probe pulse which covers a broad spectral range from 300 nm to 1400 nm. The absorption cross-section is calculated by integrating the losses of the electromagnetic field over the hybrid NP volume by exciting the system with both pump and probe pulses for different pump-probe delays. We define $\mathit{\Delta A}$ as the difference in absorption cross-section between the case of pump-probe excitation and the case of excitation by only the probe pulse (stationary state). In Fig. 5, we present the TA response at the plasmonic excitation of CuS (i.e. 1051 nm). We remark that an important role in explaining the dual-plasmonic response revealed by TA is played by the optical properties of CuS. CuS NPs exhibit "hyperbolic" behavior between 670 nm and 1050 nm, meaning that the material has metallic and dielectric permittivity for in-plane and out-of-plane polarization, respectively [8]. Modeling the anisotropic properties of CuS NSs is important especially in the transient regime where the coupling of CuS with Au NPs is modified due to hot charge carriers accumulations at the border of CuS and Au.

In Fig. 5a, we focus on the impact of Landau damping (LD) on the dual-plasmonic response. In hybrid Au/CuS NSs, the charge carriers photogenerated via the LD mechanism can accumulate at the border between Au and CuS before the thermalization process takes effect in the system. These charger carriers modify the permittivity of the NPS over a thin region, thus resulting in a modification of their resonance condition, which can be probed over a short time interval after photoexcitation by TAS. After photoexcitation of a specific plasmonic resonance, the dephasing of the LSPR gives rise to the generation of hot carriers distributed at energy states well above and below the Fermi level (see Fig. 2 Ref [29]). The temporal dynamics of the LSPR excitation is followed by the generation of hot charge carriers, which can be described by the LD mechanism associated with the electronic excitations and internal non-radiative relaxation processes of each particle [29],[16]. Within the first few hundred femtoseconds after photoexcitation, LD due to surface collisions is dominant in the system where the LSPR's decay occurs without interaction with phonons or defects [29],[16]. Over a longer time period, the system evolves from the non-thermal to a thermal relaxation process, eventually equilibrating to the steady-state [4-5] (a discussion on the thermalization process which determines the tail of the absorption response up to some tens of picoseconds is out of the scope of this paper). The Landau effect is taken into account by correcting the permittivity of Au and CuS in the pump-probe simulation via the Lindhard's formula. For simplicity, we consider this correction time-independent. The Lindhard's formula is given by [30],[16],

$$\varepsilon^{LD}(\omega, k) = \varepsilon_b + \frac{3\omega_p^2}{k^2 v_f^2}\left(1 - \frac{\omega}{2kv_F} ln\left(\frac{\omega+kv_F}{\omega-kv_F}\right)\right), \qquad (1)$$

which has an imaginary part for for $k > \omega/v_F$ [16] in addition to the real part [31]. The formula is numerically calculated at the Fermi temperature $T \approx T_F$, where $\varepsilon_b$ is the relative permittivity at high frequencies, $k = k_F$ is the Fermi wavenumber, $v_F$ is the Fermi velocity and $\omega_p$ is the plasma frequency equal to 21.81× $10^{14}$ Hz for Au [16] and equal to 10.881× $10^{14}$ Hz for CuS [32]. The correction of the permittivity due to the Lindhard's model happens at the common interface between the Au and CuS NPs, which is described as the Landau region. After generation, more than half of the generated hot charge carriers remains in the vicinity of the interparticle interface within the Landau thickness $\Delta L$, where $\Delta L = 2\pi/\Delta k = v_F/\nu$ is a function of the optical frequency $\nu$. The Landau region $\Delta L$ is schematically shown in the inset of Fig. 5a. In the investigated wavelength range, its thickness, which is related to one optical oscillation period, is much smaller than the electron/hole mean free path [16] for both Au and CuS. In fact, charge carriers generated due to the LD mechanism (also called surface collision assisted "tilted" transition), have not sufficient energy to overcome the surface (Schottky) barrier Φ between two adjacent media, hence they remain at the vicinity of the interface [16]. The values of $\varepsilon_b$= 1.53, $v_F$= 1.4× $10^8$ cm/s, $k_F = 1.20 \times 10^8$/cm were adopted in numerical simulations for gold NSs [16], while for CuS NPs, the values of $\varepsilon_b$ = 9 [32], $v_F = 7.1762\times 10^7$ cm/s [33], $k_F = 4.9565 \times 10^7$/cm [32] were used.

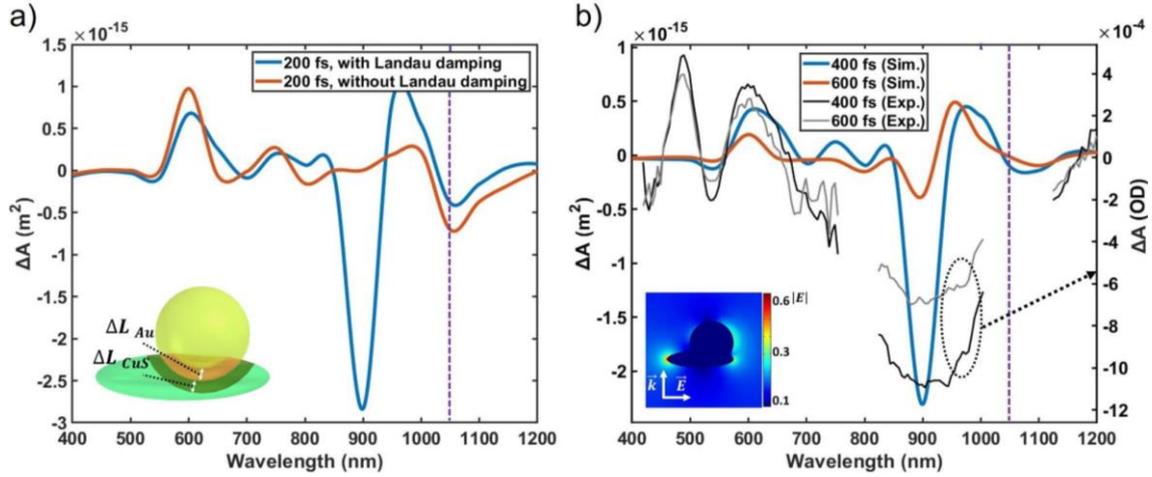

**Fig. 5**. a) Impact of Landau damping correction in the transient regime for excitation at 1051 nm and pump-probe delay of 200 fs. The inset shows the Landau regions for both materials schematically. b) Numerical vs. experimental dual-plasmonic TA response of Au/CuS UFO-shaped NSs to a pump excitation of 1051 nm; pump-probe time delays of 400 fs and 600 fs are considered (each measurement is obtained by averaging five scans over a ±40 fs time interval). The inset shows a snapshot of the electric field distribution in the transient regime for a hybrid NS pumped at 1051 nm (at t = 400 fs after photoexcitation), indicating the simultaneous plasmonic excitation in both Au and CuS. Landau damping is accounted for in simulations presented in b).

A comparison of the optical response with and without Landau damping contribution is shown in Fig. 5a for excitation at 1051 nm for a pump-probe delay of 200 fs (the case of excitation at 551 nm is shown in Fig. S5a, only numerical simulations). In general, a negative *ΔA* indicates a photoinduced bleaching of the ground-state population density (due to a promotion of electrons to the excited state) as a result of the pump excitation [28]. A negative *ΔA* dip is generally accompanied by positive peaks at longer and shorter wavelengths [28]. Introducing the Landau damping modifies the *ΔA* spectrum for both Au and CuS LSPRs. While the change is not so pronounced around the LSPR of Au (around λ = 600 nm), the change in *ΔA* is much more substantial in the IR spectral range, where the LSPR of CuS is located. We see the emergence of a dip in *ΔA* at λ = 900, which is significantly blue-shifted with respect to the steady-state LSPR optical response of CuS, usually located above ~1000 nm. This makes the TAS of the CuS-LSPR overlapping more with the transient response of the Au-LSPR. The *ΔA* spectrum obtained with the inclusion of the Landau damping mechanism allows us to qualitatively reproduce the experimental trends. This becomes evident in Fig. 5b, where the comparison between TA simulations and measurements is presented for a pump-probe delay of 400 fs and 600 fs. We observe a transient response at the spectral position of both Au and CuS LSPR bands. Furthermore, the amplitude of such peaks and dips in *ΔA* decreases for increasing pump-probe delay as expected. It is remarkable that the transient response from the Au-LSPR follows the typical pattern of the transient response of a resonantly excited LSPR (bleach with two induced absorptions at its sides), although the Au-LSPR was not excited resonantly. This is due to the plasmonic coupling with CuS. Here we would like to stress that our simulations of the TA response only qualitatively match with experimental measurements. Especially since the experimentally measured IR response is much broader than the modeled responses, no new distinct peak is visible in the measurements of the hybrid system when compared to the pure CuS system. However, the overall position of the broad measured NIR signal can be well explained with the LD associated response. Further the magnitudes of the different simulated responses in the NIR and the visible part only relatively match with the experiments (e.g. the response stemming from Au at 1051 nm excitation is always much weaker in simulations than in the measured TA spectra). Nevertheless the introduction of LD in the simulation helps to make the match between simulation and experiment much better concerning the position of the NIR response and slightly better concerning the amplitude of the Au response in the visible, making it very reasonable to assume that LD plays a crucial role in the hybrid particle response at short time-scales after excitation. The inset in Fig. 5b shows the electric field distribution under pump-probe excitation for a hybrid AuCu NS in the transient regime, indicating the simultaneous emergence of LSPRs in Au and CuS. The dip in $\Delta A$ can red- or blue-shift due to the changes in the charge carrier distribution induced by the pump excitation, leading to a transiently broadening of the LSPR absorption band [34],[35]. This, together with the distribution of NP size and geometrical configurations, already discussed in Section 2, can explain the broader response observed in experiments compared to simulations. Our numerical study remarks the role played by the generation of hot carriers in such a system and the need of introducing Landau damping in numerical modeling.

## 5. Conclusions

In this paper, we studied numerically and experimentally the dual-plasmonic response of Au/CuS UFO-shaped NSs. Our simulations clarify previously reported experimental measurements by describing the effect of NP topology, orientation, Au/CuS size ratio, and the role played by the hyperbolic anisotropic permittivity of CuS. In the steady-state regime, experimental investigations revealed that increasing the size of CuS significantly enhances the optical response in the IR regime while reducing the response in the visible. Although simulations confirm many of these trends, the simplified model of one single NS cannot explain all of them. Further investigations are therefore needed to account for the statistics of the NPs, as well as the effect of their clustering/proximity. Moreover, we proposed an approach for the modeling of TA via simulations in the time-domain under pump-probe excitation. We accounted for the hot-charge carriers' generation process at the interface between Au and CuS by introducing a correction in the permittivity based on the Landau damping mechanism, which is valid before the thermalization process becomes dominant in the system. Our simulations qualitatively reproduce the trends observed in experiments, thus providing an important tool for the modeling of TAS.

## 6. Methods

**Used Chemicals**
Chloroform (99.8%), copper(II)acetylacetonate (Cu(acac)$_2$, 99.99%), 1,2-dichlorobenzene (DCB, 99%), methanol (MeOH, 99.8%), 1-octadecene (ODE, 90%), oleylamine (OLAM, 70%), sulfur (S, 99.98%), and toluene (≥99.5%) were purchased from Sigma Aldrich. Ethanol (EtOH, 99.8%) was purchased from Roth. Hydrogen tetrachloroaurate(III) trihydrate (HAuCl$_4$·3 H$_2$O, 99.99%) was purchased from Alfa Aesar. All chemicals were used as received without further purification.

**Synthesis of Au NPs**
The Au NPs used as seeds for the synthesis of hybrid Au/CuS NPs were synthesized following a procedure of Sun et al. [36]. First 5 ml OLAM were heated to 160 °C under the Ar atmosphere while stirring. This temperature was kept for 20 min. 0.1 ml of a 1 M solution of $HAuCl_4·3\,H_2O$ in water was added swiftly. While stirring, the reaction was kept at 160 °C for 60 min and then cooled to room temperature naturally. For washing MeOH (≈10 ml) was added and the solution centrifuged for 10 min at 10621 rcf. The supernatant was discarded, the NPs were redispersed in 5 ml DCB.

**Synthesis of hybrid Au/CuS NPs**
The hybrid Au/CuS NPs used as seeds for the seed mediated growth reactions were synthesized following a procedure of Sun et al [36], 26.1 mg $Cu(acac)_2$, 1 mL OLAM and 5 mL of the previously synthesized Au NPs dispersed in DCB were heated to 80 °C while stirring. 1 mL of a 0.1 M solution of S in DCB was added swiftly and the solution was heated to 100 °C under Ar atmosphere. The temperature was kept for 30 min and cooled down to room temperature naturally. For washing EtOH (≈10 mL) was added and the reaction solution was centrifuged for 20 min at 3773 rcf. The supernatant was discarded and the NPs were dispersed in toluene (for analytic) or DCB (for seed mediated growth).

**Increasing the CuS-domain size of hybrid Au/CuS NPs via seed mediated growth**
The increase of the CuS-domain size followed a modified procedure of Sun et al. [36]. The procedure is the same as for growth of CuS-domains on Au NPs as described above, with the exception that instead of pure Au NPs in DCB hybrid Au/CuS NPs in DCB were used as seeds. The size of the CuS-domains can further be increased by iteratively repeating the synthesis with the obtained Au/CuS NPs.

**Transmission electron microscopy**
TEM images were obtained using a FEI Tecnai G2 F20 device equipped with a field emission gun. The acceleration voltage was 200 kV. The samples were washed by precipitation with EtOH and centrifugation. The nanocrystals were dispersed in chloroform and 20 µL of this dispersion were dropcasted on a 300 mesh Cu grid by Quantifoil.

**UV/Vis/Near-IR-spectroscopy**
UV/Vis/Near-IR spectra were collected using a Cary 5000 spectrophotometer equipped with an external integrating sphere (DRA2500) by Agilent Technologies. The samples were diluted in toluene and placed in the center position of the integrating sphere in quartz glass cuvettes with 10 mm path length. A gratin and detector change was conducted at 800 nm and a lamp change was conducted at 350 nm.

**TAS**
Samples were prepared as colloidal solutions in toluene in quartz glass cuvettes with a 2 mm pathway and were constantly stirred in order to prevent laser mediated phase transitions or decomposition [37],[38]. Ultrafast charge carrier dynamics were studied by broadband pump-probe spectroscopy in a set-up described previously and briefly discussed here [39],[40],[41]; 120 fs laser pulses with a wavelength of 805 nm were generated by a Ti:sapphire laser and split in order to get pump and probe pulses. The pump beam wavelength was adjustable via nonlinear frequency mixing in an optical parametric amplifier and second harmonics generation system (TOPAS Prime from Spectra-Physics). The probe pulse was converted to a broadband supercontinuum using nonlinear processes in a $CaF_2$ or sapphire crystal. The probe pulse could be delayed up to 8 ns after photoexcitation via an automatic delay line. Pump and probe pulse pass the cuvette with a pathlength of 2 mm and the maximum overlap. After transmission, the probe pulse was measured by a fiber-coupled detector array at different delay times to follow ultrafast processes causing changes in the absorption.

**Acknowledgements**
We acknowledge the Deutsche Forschungsgemeinschaft (DFG, German Research Foundation) under Germany's Excellence Strategy within the Cluster of Excellence PhoenixD (EXC 2122, Project ID 390833453) and the Cluster of Excellence CUI: Advanced Imaging of Matter´ (EXC2056, project ID 390715994). We acknowledge the Leibniz Young Investigator Grants program by the Leibniz University Hannover (Grant Ref. Num: LYIG-2023-04). The TA measurements were funded by the DFG under contract INST 37/1160-1 FUGG (project nr. 458406921). A. A. and J. L. thankfully acknowledge additional funding by the MWK Niedersachsen. P.B. is grateful for being funded by the Hannover School for Nanotechnology (HSN). We acknowledge the computing time granted by the Resource Allocation Board and provided on the supercomputer Lise and Emmy at NHR@ZIB and NHR@Göttingen as part of the NHR infrastructure (project nip00059).

**References**
[1] A. Ganai, P. S. Maiti, L. Houben, R. Bar-Ziv, and M. Bar Sadan, "Inside-Out: The Role of Buried Interfaces in Hybrid Cu2ZnSnS4–Noble Metal Photocatalysts," *J. Phys. Chem. C*, vol. 121, no. 12, pp. 7062–7068, 2017, doi: 10.1021/acs.jpcc.7b01733.
[2] Y. Kim *et al.*, "Synthesis of Au−Cu2S Core−Shell Nanocrystals and Their Photocatalytic and Electrocatalytic Activity," *J. Phys. Chem. C*, vol. 114, no. 50, pp. 22141–22146, 2010, doi: 10.1021/jp109127m.
[3] B. Li *et al.*, "Cu 7.2 S 4 nanocrystals: a novel photothermal agent with a 56.7% photothermal conversion efficiency for photothermal therapy of cancer cells," *Nanoscale*, vol. 6, no. 6, pp. 3274–3282, 2014, doi: 10.1039/C3NR06242B.
[4] S. K. Ghosh and T. Pal, "Interparticle Coupling Effect on the Surface Plasmon Resonance of Gold Nanoparticles:  From Theory to Applications," *Chem. Rev.*, vol. 107, no. 11, pp. 4797–4862, 2007, doi: 10.1021/cr0680282.
[5] H. Lee, S. W. Yoon, E. J. Kim, and J. Park, "In-Situ Growth of Copper Sulfide Nanocrystals on Multiwalled Carbon Nanotubes and Their Application as Novel Solar Cell and Amperometric Glucose Sensor Materials," *Nano Lett.*, vol. 7, no. 3, pp. 778–784, 2007, doi: 10.1021/nl0630539.
[6] M. H. Rahaman and B. A. Kemp, "Analytical model of plasmonic resonance from multiple core-shell nanoparticles," *Opt. Eng.*, vol. 56, no. 12, p. 121903, 2017, doi: 10.1117/1.OE.56.12.121903.
[7] C. Hu *et al.*, "Generating plasmonic heterostructures by cation exchange and redox reactions of covellite CuS nanocrystals with Au 3+ ions," *Nanoscale*, vol. 10, no. 6, pp. 2781–2789, 2018, doi: 10.1039/C7NR07283J.
[8] R. M. Córdova-Castro *et al.*, "Anisotropic Plasmonic CuS Nanocrystals as a Natural Electronic Material with Hyperbolic Optical Dispersion," *ACS Nano*, vol. 13, no. 6, pp. 6550–6560, 2019, doi: 10.1021/acsnano.9b00282.
[9] Q. Feng *et al.*, "A smart off–on copper sulfide photoacoustic imaging agent based on amorphous–crystalline transition for cancer imaging," *Chem. Commun.*, vol. 54, no. 78, pp. 10962–10965, 2018, doi: 10.1039/C8CC06736H.
[10] G. Ku, M. Zhou, S. Song, Q. Huang, J. Hazle, and C. Li, "Copper Sulfide Nanoparticles

As a New Class of Photoacoustic Contrast Agent for Deep Tissue Imaging at 1064 nm," *ACS Nano*, vol. 6, no. 8, pp. 7489–7496, 2012, doi: 10.1021/nn302782y.

[11] J. Mou *et al.*, "Ultrasmall Cu2-xS Nanodots for Highly Efficient Photoacoustic Imaging-Guided Photothermal Therapy," *Small*, vol. 11, no. 19, pp. 2275–2283, 2015, doi: 10.1002/smll.201403249.

[12] H. Zhu *et al.*, "Monodisperse Dual Plasmonic Au@Cu2–xE (E= S, Se) Core@Shell Supraparticles: Aqueous Fabrication, Multimodal Imaging, and Tumor Therapy at in Vivo Level," *ACS Nano*, vol. 11, no. 8, pp. 8273–8281, 2017, doi: 10.1021/acsnano.7b03369.

[13] B. Shan, Y. Zhao, Y. Li, H. Wang, R. Chen, and M. Li, "High-Quality Dual-Plasmonic Au@Cu2–xSe Nanocrescents with Precise Cu2–xSe Domain Size Control and Tunable Optical Properties in the Second Near-Infrared Biowindow," *Chem. Mater.*, vol. 31, no. 23, pp. 9875–9886, 2019, doi: 10.1021/acs.chemmater.9b04100.

[14] L. Chen, H. Hu, Y. Chen, J. Gao, and G. Li, "Plasmonic Cu $_{2-x}$ S nanoparticles: a brief introduction of optical properties and applications," *Mater. Adv.*, vol. 2, no. 3, pp. 907–926, 2021, doi: 10.1039/D0MA00837K.

[15] P. Bessel, A. Niebur, D. Kranz, J. Lauth, and D. Dorfs, "Probing Bidirectional Plasmon-Plasmon Coupling-Induced Hot Charge Carriers in Dual Plasmonic Au/CuS Nanocrystals," *Small*, vol. 19, no. 12, p. 2206379, 2023, doi: 10.1002/smll.202206379.

[16] J. Khurgin, A. Y. Bykov, and A. V. Zayats, "Hot-electron dynamics in plasmonic nanostructures", 2023, arXiv preprint arXiv:2302.10247.

[17] E. Prodan, C. Radloff, N. J. Halas, and P. Nordlander, "A Hybridization Model for the Plasmon Response of Complex Nanostructures," *Science*, vol. 302, no. 5644, pp. 419–422, 2003, doi: 10.1126/science.1089171.

[18] Y. Xie *et al.*, "Copper Sulfide Nanocrystals with Tunable Composition by Reduction of Covellite Nanocrystals with Cu+ Ions," *J. Am. Chem. Soc.*, vol. 135, no. 46, pp. 17630–17637, 2013, doi: 10.1021/ja409754v.

[19] H. Bao *et al.*, "Ultrathin and Isotropic Metal Sulfide Wrapping on Plasmonic Metal Nanoparticles for Surface Enhanced Raman Scattering-Based Detection of Trace Heavy-Metal Ions," *ACS Appl. Mater. Interfaces*, vol. 11, no. 31, pp. 28145–28153, 2019, doi: 10.1021/acsami.9b05878.

[20] S. A. Maier, *Plasmonics: Fundamentals and Applications*. New York, NY: Springer US, 2007. doi: 10.1007/0-387-37825-1.

[21] C. F. Bohren and D. R. Huffman, *Absorption and Scattering of Light by Small Particles*. John Wiley & Sons, 2008.

[22] A. Trügler, *Optical Properties of Metallic Nanoparticles: Basic Principles and Simulation*. Springer, 2016.

[23] G. Ritzoulis, N. Papadopoulos, and D. Jannakoudakis, "Densities, viscosities, and dielectric constants of acetonitrile + toluene at 15, 25, and 35 .degree.C," *J. Chem. Eng. Data*, vol. 31, no. 2, pp. 146–148, 1986, doi: 10.1021/je00044a004.

[24] L. Xiao *et al.*, "Near-infrared radiation absorption properties of covellite (CuS) using first-principles calculations," *AIP Adv.*, vol. 6, no. 8, p. 085122, Aug. 2016, doi: 10.1063/1.4962299.

[25] A. L. Routzahn, S. L. White, L.-K. Fong, and P. K. Jain, "Plasmonics with Doped Quantum Dots," *Isr. J. Chem.*, vol. 52, no. 11–12, pp. 983–991, 2012, doi: 10.1002/ijch.201200069.

[26] D. Dorfs *et al.*, "Reversible Tunability of the Near-Infrared Valence Band Plasmon Resonance in Cu $_{2-x}$ Se Nanocrystals," *J. Am. Chem. Soc.*, vol. 133, no. 29, pp. 11175–11180, 2011, doi: 10.1021/ja2016284.

[27] J. M. Luther, P. K. Jain, T. Ewers, and A. P. Alivisatos, "Localized surface plasmon resonances arising from free carriers in doped quantum dots," *Nat. Mater.*, vol. 10, no. 5, Art. no. 5, 2011, doi: 10.1038/nmat3004.

[28] J. Lauth, S. Kinge, and L. D. A. Siebbeles, "Ultrafast Transient Absorption and Terahertz Spectroscopy as Tools to Probe Photoexcited States and Dynamics in Colloidal 2D Nanostructures," *Z. Für Phys. Chem.*, vol. 231, no. 1, pp. 107–119, 2017, doi: 10.1515/zpch-2016-0911.

[29] A. Schirato, M. Maiuri, G. Cerullo, and G. D. Valle, "Ultrafast hot electron dynamics in plasmonic nanostructures: experiments, modelling, design," *Nanophotonics*, vol. 12, no. 1, pp.

1–28, 2023, doi: 10.1515/nanoph-2022-0592.

[30] J. Lindhard, "ON THE PROPERTIES OF A GAS OF CHARGED PARTICLES," *Kgl Dan. Vidensk. Selsk. Mat-Fys Medd*, vol. Vol: 28, No. 8, 1954.

[31] A. V. Andrade-Neto, "Dielectric function for free electron gas: comparison between Drude and Lindhard models," *Rev. Bras. Ensino Física*, vol. 39, no. 2, 2016, doi: 10.1590/1806-9126-rbef-2016-0206.

[32] Y. Xie *et al.*, "Metallic-like Stoichiometric Copper Sulfide Nanocrystals: Phase- and Shape-Selective Synthesis, Near-Infrared Surface Plasmon Resonance Properties, and Their Modeling," *ACS Nano*, vol. 7, no. 8, pp. 7352–7369, 2013, doi: 10.1021/nn403035s.

[33] J. P. Tailor, S. H. Chaki, and M. P. Deshpande, "Comparative study between pure and manganese doped copper sulphide (CuS) nanoparticles," *Nano Express*, vol. 2, no. 1, p. 010011, 2021, doi: 10.1088/2632-959X/abdc0d.

[34] T. S. Ahmadi, S. L. Logunov, and M. A. El-Sayed, "Picosecond Dynamics of Colloidal Gold Nanoparticles," *J. Phys. Chem.*, vol. 100, no. 20, pp. 8053–8056, 1996, doi: 10.1021/jp960484e.

[35] S. Link and M. A. El-Sayed, "Shape and size dependence of radiative, non-radiative and photothermal properties of gold nanocrystals," *Int. Rev. Phys. Chem.*, vol. 19, no. 3, pp. 409–453, 2000, doi: 10.1080/01442350050034180.

[36] C. Sun, M. Liu, Y. Zou, J. Wei, and J. Jiang, "Synthesis of plasmonic Au–CuS hybrid nanocrystals for photothermal transduction and chemical transformations," *RSC Adv.*, vol. 6, no. 31, pp. 26374–26379, 2016, doi: 10.1039/C6RA02425D.

[37] D. Kranz *et al.*, "Size-Dependent Threshold of the Laser-Induced Phase Transition of Colloidally Dispersed Copper Oxide Nanoparticles," *J. Phys. Chem. C*, vol. 126, no. 36, pp. 15263–15273, 2022, doi: 10.1021/acs.jpcc.2c03815.

[38] M. Niemeyer *et al.*, "Nanosecond Pulsed Laser-Heated Nanocrystals Inside a Metal-Organic Framework Matrix," *Chem. Nan.o Mat*, vol. 8, no. 6, p. e202200169, 2022, doi: 10.1002/cnma.202200169.

[39] F. C. M. Spoor *et al.*, "Hole Cooling Is Much Faster than Electron Cooling in PbSe Quantum Dots," *ACS Nano*, vol. 10, no. 1, pp. 695–703, 2016, doi: 10.1021/acsnano.5b05731.

[40] A. Spreinat *et al.*, "Quantum Defects in Fluorescent Carbon Nanotubes for Sensing and Mechanistic Studies," *J. Phys. Chem. C*, vol. 125, no. 33, pp. 18341–18351, 2021, doi: 10.1021/acs.jpcc.1c05432.

[41] L. F. Klepzig *et al.*, "Colloidal 2D PbSe nanoplatelets with efficient emission reaching the telecom O-, E- and S-band," *Nanoscale Adv.*, vol. 4, no. 2, pp. 590–599, 2022, doi: 10.1039/D1NA00704A.

# Supporting Information

**Dynamic light scattering**

Dynamic light scattering (DLS) measurements were done using a Zetasizer ZSP from Malvern. The samples were diluted in toluene and placed in quartz glass cuvettes with 10 mm path length. For every dispersion 5 measurements with 10 runs every 10 s were performed. DLS investigations indicate that the particles are definitely not agglomerated in solution but are always dispersed as single particles.

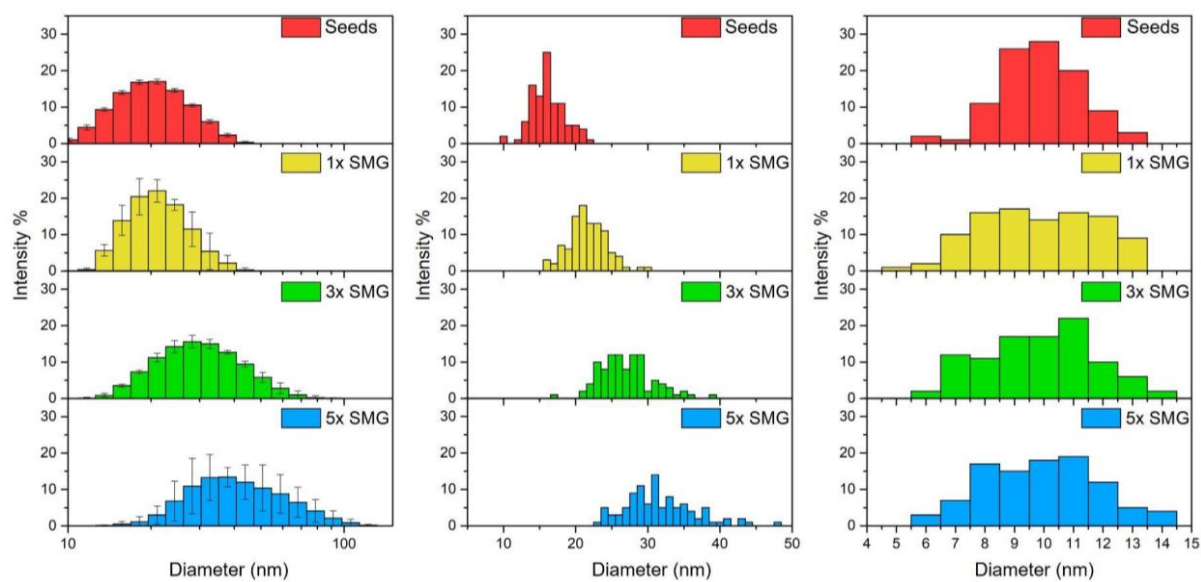

**Fig. S1.** Size distribution of the hybrid Au/CuS NPs measured with dynamic light scattering (a-d) and determined by measuring 100 NPs at their broadest point on TEM images (e-h). Size distribution of the Au-domain of the hybrid Au/CuS NPs determined by measuring 100 NPs at their broadest point on TEM images (i-l). (a,e,i) show the data for the Au/CuS NPs used as seeds in the SMG reactions. The size distributions for the NPs after one, three and five seed mediated growth steps are shown in (b,f,j), (c,g,k) and (d,h,l) respectively. Further statistical data is provided in Table S2.

| Nanoparticles | Hydrodynamic Diameter /nm | Diameter of the whole particle /nm | Diameter of the Au-domain /nm |
|---|---|---|---|
| Seeds | 13.7 ± 0.4 | 15.6 ± 2.3 | 9.4 ± 1.5 |
| 1x SMG | 16.2 ± 0.3 | 21.2 ± 2.6 | 9.4 ± 1.9 |
| 3x SMG | 18.4 ± 0.9 | 26.5 ± 3.7 | 9.4 ± 1.9 |
| 5x SMG | 24.7 ± 6.8 | 21.1 ± 4.9 | 9.4 ± 1.9 |

**Table. S1.** Hydrodynamic diameter (measured with DLS) and diameter of the whole hybrid Au/CuS NP and their Au-domain (determined by measuring 100 NPs on TEM images at their widest part).

**Additional simulation results:**

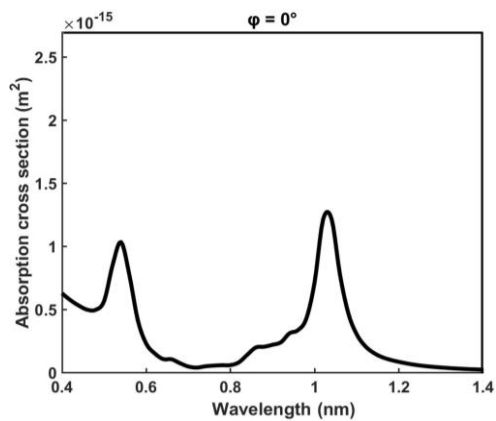

**Fig. S3.** Averaged absorption cross section spectrum of the Au/CuS UFO-shaped NS from all optical responses obtained in the range of θ = [-π/2 , π/2] with a step of 5 degree (presented in Fig 2b).

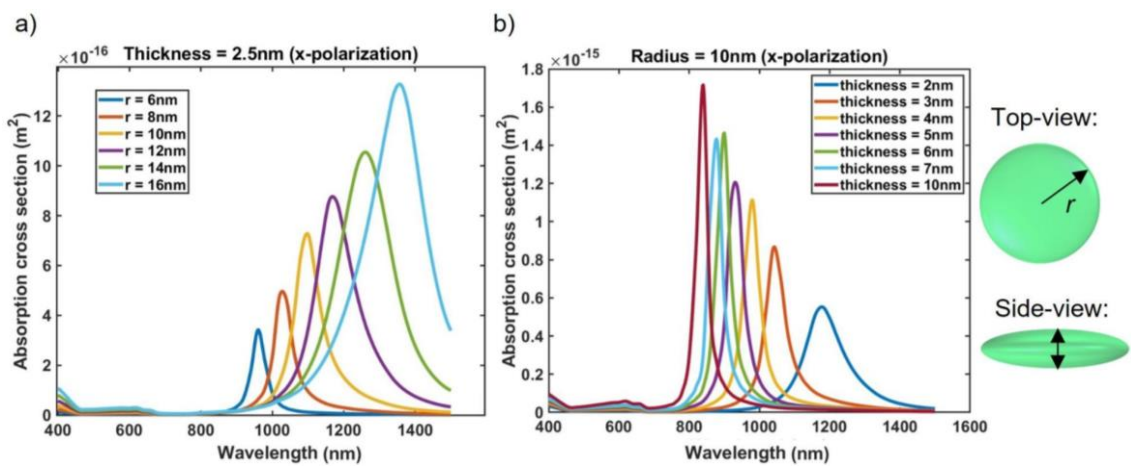

**Fig. S4.** Absorption cross section spectrum of the CuS NP as a function of the a) cross-sectional radius, and b) thickness, illuminated with a light beam polarized along the *x*-direction with propagation direction along the *z*-axis.

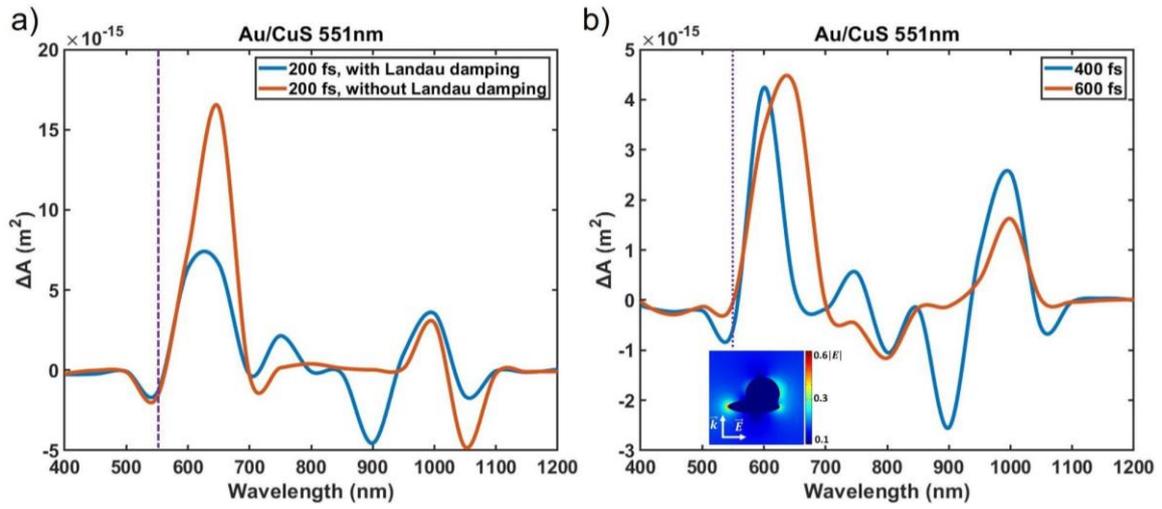

**Fig. S5**. a) Impact of Landau damping correction in the transient regime for a pump excitation at 551 nm, and for a pump-probe delay of 200 fs. b) Numerical TA response for a pump excitation at 551 nm, monitored at pump-probe delays of 400 fs and 600 fs. Inset in b) shows a snapshot of the electric field distribution in the transient regime indicating the simultaneous plasmonic excitation in both Au and CuS in hybrid Au/CuS NSs for excitation at 551 nm. Landau damping is accounted for in b).